\documentclass[a4paper,11pt]{article}
\usepackage{jinstpub} 
\usepackage{lineno}
    \usepackage{hyperref}
\usepackage{subcaption}
\newcommand{\sparc}{SPARC\_LAB }
\newcommand{\um}{$\textmu $\mathrm{m}}
\newcommand{\uJ}{\textmu J}
\newcommand{\mm}{\mathrm{mm}}
\newcommand{\nm}{\mathrm{nm}}
\newcommand{\cm}{\mathrm{cm}}
\newcommand{\picom}{\mathrm{pm}}

\title{\boldmath Mechanical strength investigations of the APPLE-X undulator using Fiber Bragg Grating strain measurements}

\author[a]{I. Balossino\note{Corresponding author.},}
\author[b]{A. Polimadei,}
\author[a]{M. Del Franco,}
\author[a]{A. Selce,}
\author[a]{A. Vannozzi,}
\author[a]{E. Di Pasquale,}
\author[a]{L. Giannessi,}
\author[b]{F. Nguyen,}
\author[b]{A. Petralia,}
\author[d]{J. Pockar,}
\author[d]{U. Primozic,}
\author[c]{R. Geometrante,}
\author[b]{M.A. Caponero,}
\author[a]{L. Sabbatini}
\affiliation[a]{INFN-National Laboratory of Frascati,\\ via Enrico Fermi 54, 00044 Frascati (Roma), Italy}
\affiliation[b]{ENEA Frascati Research Centre,\\
via Enrico Fermi 44, 00044 Frascati (Roma), Italy}
\affiliation[c]{Kyma S.p.A.\\ S.S. 14 – km 163,5 in AREA Science Park,  IT-34149 Trieste, Italy}
\affiliation[d]{Kyma tehnologija d.o.o.\\ Kraška ulica 2, SI-6210 Sežana, Slovenia}

\emailAdd{ilaria.balossino@lnf.infn.it}

\abstract{The SPARC$\_$LAB (Sources for Plasma Accelerators and Radiation Compton with Lasers and Beams) facility at the INFN (National Institute of Nuclear Physics) laboratory in Frascati is being upgraded to accommodate a new user facility as part of the SABINA (Source of Advanced Beam Imaging for Novel Applications) project. A new beamline dedicated to SABINA will be equipped with three APPLE-X undulators to deliver IR/THz radiation with photon pulses in the ps range, with energy of tens of \textmu J, and with the possibility of choosing between linear, circular, or elliptical polarization. The APPLE-X, designed and built by KYMA S.p.a., guarantees the possibility to vary the gap amplitude between the magnet’s arrays and their relative phase. The entire system, from mechanics to the kinematic subsystems, has been designed from scratch, and a structural analysis has been carried out to evaluate the production.
The undulators were delivered to Frascati in 2023 and, in collaboration with ENEA, a further investigation campaign was launched on the mechanical parts in the vicinity of the permanent magnets, using strain measurements based on optical methods. Fiber Bragg Grating (FBG) sensors consist of a phase grating inscribed in the core of a single-mode fiber, whose Bragg-diffracted light propagates back along the fiber. If bonded to the mechanical structure, they can be used as strain sensors. By following the variations in the scattered spectrum, it is possible to perform strain measurements. Using multiple FBGs applied at selected locations on the undulator, several measurements were made by following the different possible kinematics, but also by studying the quiescent response as a function of the ambient temperature. The results show a clear deformation of the structure related to the temperature changes and magnetic forces, but its magnitude is compatible or lower with respect to the one calculated with the finite elements methods and, moreover, they are well within the tolerances required for the functionality of the undulator. The tests, therefore, confirm the reliability of the mechanical structure.
}


\begin{document}
\maketitle
\flushbottom

\section{SABINA}
SABINA~\cite{Sabbatini:2021bmm} is a project of the INFN National Laboratory of Frascati (LNF), co-funded by the Lazio Region, whose goal is to create two user facilities by consolidating the existing \sparc~\cite{Ferrario:2013nka} infrastructure. An irradiation facility with a laser target line for vacuum optical samples for aerospace and, the one of interest for this article, a THz/MIR radiation line for carrying out non-invasive studies in various fields of research. 
The \sparc LINAC provides an electron beam with high-brightness and tunable energies between $30$ and $100\,\mathrm{MeV}$. For the THz line, the beam is directed into a dedicated dogleg. Here, 3 APPLE-X undulators~\cite{počkar:fel2022-wep47} operate in self-amplified spontaneous emission (SASE) mode creating photon pulses with frequency in the range $3 \div 30\,\mathrm{THz}$ or $1 \div 10\,\um$, with variable polarization, picosecond duration and energy of tens of \uJ.
The THz user facility will be promoted among scientific and industrial disciplines. The non-invasive technique that such radiation line offers is usable in several fields for studying complex phenomena and investigating inhomogeneities on a millimeter and sub-millimeter scale. It is expected to find collaborators among automotive, electronics, pharmaceutical or archeometry. 

The undulators were manufactured by Kyma S.p.a.~\cite{KYMA:1999}. They are APPLE-X type with an innovative and compact design. A successful structural analysis was performed on the critical elements, and when they arrived at LNF, an additional campaign of measurements was conducted to test potential mechanical deformation at different magnetic field values. The studies of interest in this document are carried out in collaboration with ENEA~\cite{ENEA}, using FBG instrumentation, which performs strain measurement. An overview of the experimental setup will be provided, featuring the combination of these two technologies for the first time, followed by a description of the tests done to collect data under different conditions, to have a clearer and less ambiguous perspective, as FBG sensors are sensitive to deformations caused by both temperature and mechanical stress.

The FBG~\cite{GRATTAN200040} consist of a diffraction grating inscribed periodically in an optical fiber. The reflected or transmitted spectrum of light passing through the fiber is determined by the selected period of such gratings. As the grating changes, the characteristic wavelength of the sensors shifts, providing clear information about the strain that has occurred. The ENEA Fiber Optic Sensors Laboratory provided 1$\,$cm long Broptics OS 1500 Optical Sensors, that have been readout by a Micro Optics FBG instrument. The wavelength is therefore registered at a frequency of 50Hz and the wavelength peak value is registered. The relative shift of the wavelength from the original position is the quantity that is studied and that gives the indication that a deformation is occuring. 

\begin{figure}[!h]
  \centering
  \includegraphics[width=.6\columnwidth]{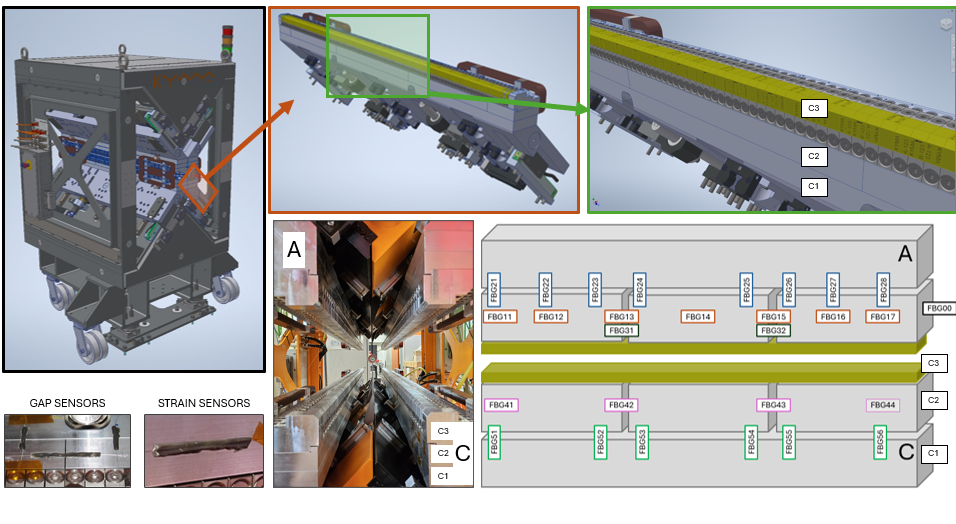}
\caption{Mechanical parts under investigation. Top part: CAD illustration of the details of the magnetic arrays; Bottom left: pictures of the real setup of magnetic arrays, gap and strain sensors; Bottom right: sensors' installation scheme to identify the sensors while in operation; }
\label{fig:1}
\end{figure}

The undulators, illustrated and schematized in figure~\ref{fig:1}, are designed to have four magnetic arrays and a gap drive mechanism. This allows to change the gap (intensity of the magnetic field) by moving all the arrays simultaneously or the phase (the polarity of the radiation) by moving longitudinally two arrays arranged diagonally (\textit{A} and \textit{C}). The gap range is $[5,150]\,\mm$; the phase shift range is $[-\frac{\lambda}{2},\frac{\lambda}{2}]$. The mechanical parts of interest in this analysis are the supports of the permanent magnet (C3) composed by three aluminum plates (C2) right below them and the single interface plate (C1) towards the girder.

The sensors were installed on the arrays that are allowed to move also longitudinally, made the assumption that they would be the most stressed ones.
28 sensors (see scheme in bottom right of figure~\ref{fig:1}) were glued with Araldite 2014, chosen following several past experiences from ENEA. The naming is chosen to group sensors with the same orientation with respect to the longitudinal axis as well as they installation order. They allow the investigation of three different behaviors: 
\begin{itemize} \setlength\itemsep{-0.5em}\labelsep=1mm \itemindent=-6mm %
    \item \textbf{environment}: 1$\,$FBG monitors the behaviour with respect to the environment's temperature. It is on a face of the metal plate outside the APPLE-X (FBG00 in figure~\ref{fig:1}), close to the temperature sensors
    \item \textbf{gap sensors}: 20$\,$FBG placed on the surfaces facing the inside of the device between adjacent plates, either in vertical or horizontal position. In this case the strain measure gives indication of the changes in the separation space, overestimated to be at maximum $1\,\mm$
    \item \textbf{strain sensors}: 7$\,$FBG fully adhesive to the surface. The measurement in this case would be related to a mechanical deformation of the plate itself.
\end{itemize}

\section{Measurements}

The measurements are made by following the peak of the spectrum with respect to the value at the beginning of the test: it is a comparative test. The value of the wavelength, $\mathrm{\lambda}\,[\nm]$ is then converted into a strain parameter following the rule that $\Delta\mathrm{\lambda}=1.2\,\picom$  corresponds to $1\,$\textmu $\mathrm{strain}$, defined as well as a deformation of $1\,\um / \mathrm{m} $. It is important to emphasize that in the following paragraphs the measurements are presented as the relative deformation that occurs to the sensors while moving the undulator mechanical parts with respect to the position that they have at the beginning of the test. 

\begin{figure}[h]
   \centering
   \includegraphics*[width=0.9\columnwidth]{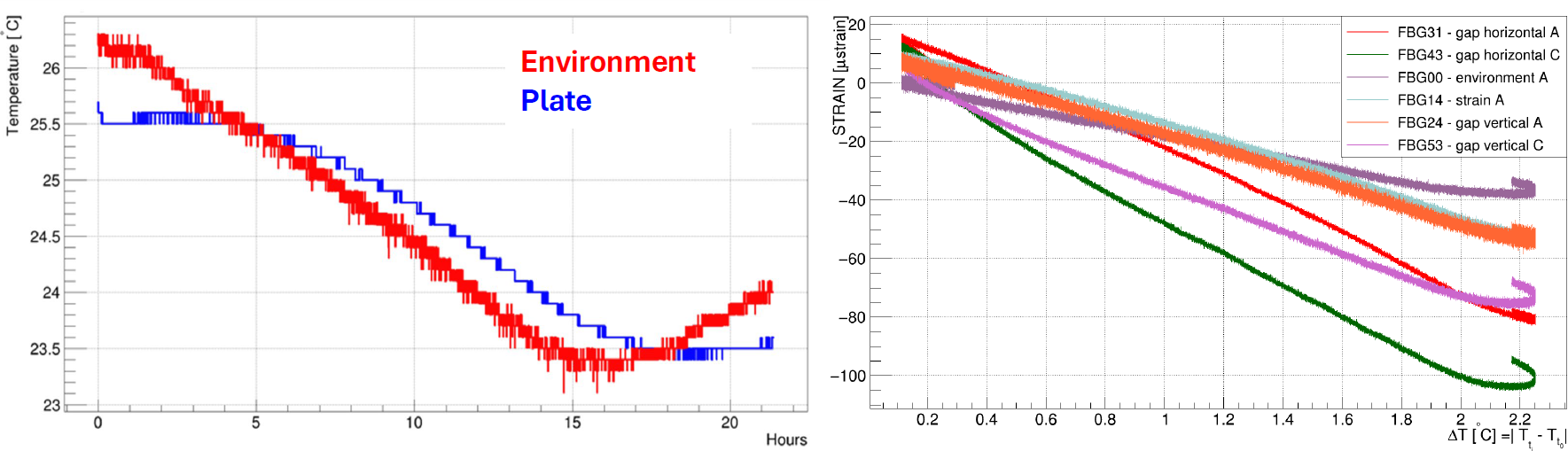}
   \caption{\centering Left: Temperature vs Time; Right: Strain vs $\Delta T$ for the most significative FBG sensors}
   \label{fig:ENV}
\end{figure}

The response to temperature variation is tested with a static acquisition over 22 hours. Two temperature's sensors are used to give both a measure of the environment around the undulator (\textit{environment}) and a measure of the metal plate, with the sensor installed adherent to the metal plate (\textit{plate}). The latter is positioned on the undulator in the same way as the FBG sensors. Therefore, it is the one that will be taken into account in the following considerations. Its $\Delta$T was measured to be about {$2.2\, ^{\circ}\mathrm{C}$}. The right plot of figure~\ref{fig:ENV} shows the strain measured during this long acquisition as a function of the temperature variation of the plate sensor. The most significant sensors were chosen to be shown: from each magnetic array, A and C, one horizontal (FBG31, FBG43) and one vertical (FBG24, FBG53) gap sensors, as well as, a strain sensor (FBG14) and the environment (FBG00) one. This test was helpful to notice that the gap sensors are the one suffering a more significant change, while the strain sensors show changes compatibles to the results obtained with the environmental sensor, sensitive only to the thermal expansion of the metal.

\begin{figure}[h]
   \centering
   \includegraphics*[width=0.9\columnwidth]{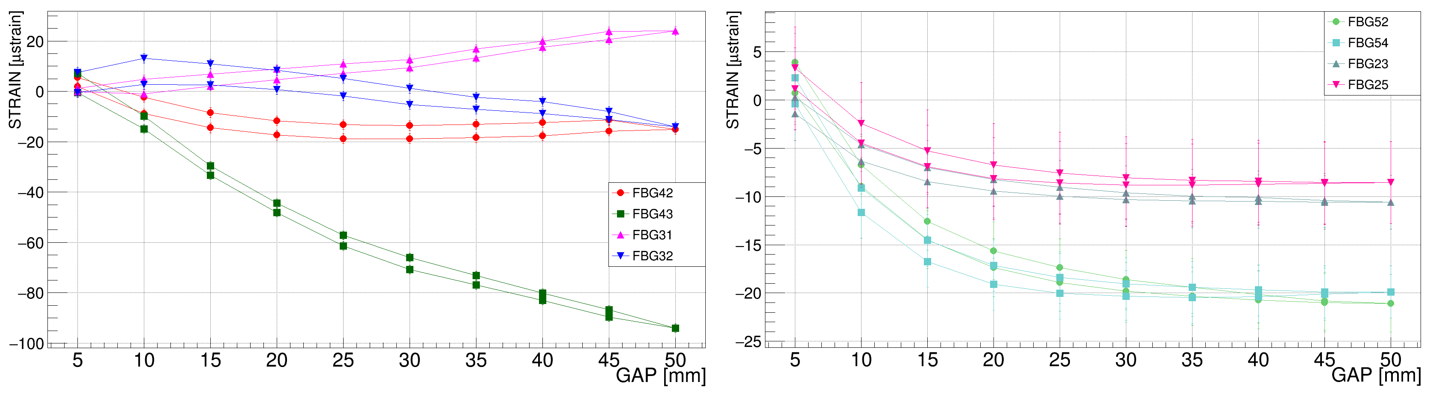}
   \caption{Strain vs Gap Amplitude. Left: Horizontal gap sensors; Right: Vertical gap sensors.}
   \label{fig:OPEN}
\end{figure}

The test to study the effects of the different intensity of the magnetic field is done registering the data while changing the gap of the undulator. Here, figure~\ref{fig:OPEN} presents the test performed moving from the minimum gap amplitude up to $\mathrm{50 \,mm}$ and back, with $\mathrm{5\,mm}$ steps; on the left the horizontal sensors, on the right the vertical sensors. It reports only the most relevant behavior for the discussion. Each gap shows two points corresponding to the aperture and closure phases: the small hysteresis is due to the temperature change during the whole test that lasted about $\mathrm{45\,minutes}$. In both plots it is possible to notice a sort of change of trend around gap amplitudes of about $\sim\mathrm{15}/ \mathrm {20\,mm}$. For the horizontal sensors, at small gaps, different magnetic arrays (A and C) behave in a opposite way, compatible with an effect due to the magnetic force. 

After that gap, the magnetic force decrease and the behavior of the sensors changes: probably a movement of the central plate due to the combined effect of magnetic, gravitational and mechanical forces it is the cause that sensors around the same plate are either stretched or compressed. The vertical sensors show an increase of the stress up until about $\mathrm {20\,mm}$ probably due to the magnetic force. After that point the strain remains stable and recovers on the way back to the original position.

\begin{figure}[!htb]
   \centering
   \includegraphics*[width=0.9\columnwidth]{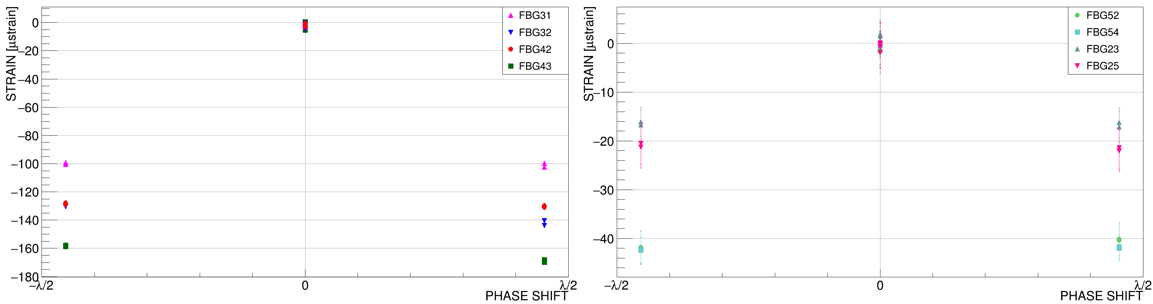}
   \caption{Strain vs Phase Shift. Left: Horizontal gap sensors; Right: Vertical gap sensors.}
   \label{fig:SHIFT}
\end{figure}

This last test is performed to investigate the last possible movements that the undulator can do: phase shift. Figure~\ref{fig:SHIFT} shows the most relevant steps of the tests performed at the minimum gap amplitude, i.e. at the higher magnetic force.  The strain is worse for the horizontal sensors (left plot) with respect to the vertical ones (right plot). Comparing it with the results previously shown, the horizontal sensors present an overall deformation of hundred \textmu m/m (wavelength shift from the original position of hundreds of pm), way lower than a stress that could affect the undulator performances. It is also relevant also that the results are reproducible and that shifts in either direction causes the same stresses. 

\section{Conclusions}

The described tests are performed with a well-known technology such as FBG sensors applied in a new way to investigate the performance of a new mechanical structure such the one of the undulator as presented also in~\cite{IPAC24}. 
The results were important to confirm that the high magnetic field does not affect the mechanical structure. They highlighted overall the extreme sensitivity of this measurement system to both temperature changes and mechanical movements and stress.
In detail, a distinction can be done between the gap and the strain sensors, while keeping in mind that all the measurements presented are relative deformation from the situation at the beginning of the test. The greatest deformation occurs for the former, which is at most $\sim\mathrm{180\,}$\textmu strain, among all the tests. For them, it is considered the length of the sensor not adherent to the metal by over-estimating the space between the metal plates to be $\sim\mathrm{1\mm}$. The results show a variation of the gap itself of $\sim\mathrm{180\,\nm}$.
The strain sensors show a deformation lower of a factor 2 to 5. Here, the deformation occurs over the full $\mathrm{1\,\cm}$ length of the sensors and can be estimated to be at worst of $\sim\mathrm{330\,\nm}$.
 The deformations are compatible or lower than the ones calculated by the finite elements method \cite{počkar:fel2022-wep47} and, therefore, confirm the reliability of the undulator mechanics.

\acknowledgments

This work is supported by NextGeneration EU- Italian National Recovery and Resilience Plan, Mission 4 - Component 2 - Investment 3.1. - Project name: Rome Technopole, CUP: I93C21000150006. SABINA is a project co-funded by Regione Lazio with the “Research infrastructures” public call within PORFESR 2014-2020.


\end{document}